# Random Access Concatenated Libraries and *dd* enable a short-latency, content-rich website on an inexpensive shared server

Don Krieger

## Abstract

Content-rich websites typically house their images as individual files or as more costly binary database objects. Both methods place high demands on storage resources with commensurate high monetary cost. Inexpensive shared service accounts come with strict limits on disk and computing resources which prevent their use for hosting content-rich websites. To minimize the load on these limited resources, large numbers of images or other records may be concatenated into a single file and delivered with short latency independent of location within the file using Linux utility dd. This solution and its performance are presented for (a) a website which houses 8000+ genealogical reference books with more than 3.5 million 250-Kbyte page images, (b) a website-based database of 86 million death records, and (c) a cluster computing application which utilizes 800 neuroimaging databases, each with 1.5 million records. These disparate applications demonstrate the efficacy, scalability, and general utility of the solution.

# Introduction

Efficient means for website storage and subsequent delivery of graphic images is problematic when the image count is high. This is so because content-rich websites typically house their images either as individual files or as more costly binary database objects, both of which place high demands on storage resources with commensurate high monetary cost. These costs are reflected in file count limits imposed by hosting providers for inexpensive shared accounts, and storage space limits for more costly virtual private and dedicated server accounts [1,2]. Cloud providers use this latter cost per unit storage space model as well [3-5]. While shared accounts are offered for under $10/month by many providers, websites which host more than 200,000 images almost always require resources only provided for accounts costing $100's/month at minimum.

The work reported here using Random Access Concatenated Libraries (RacLibs) for image storage originated as a solution to limitations imposed by a shared web hosting service on a website designed to provide rapid access to more than 8000 genealogical reference volumes with total page count in excess of 3.5 million. At current prices, hosting for the site using standard image storage methods would be at least $1000/month. At launch in 2010 the cost would have been much higher.

The shared server solution uses standard CGI/shell-callable tools provided with the Unix and Linux operating systems. RacLib files as large as 13 GBytes, each containing 1000's of jpeg images, are handled as random access memory using utility *dd*. Accessing individual images within large files using standard serial archive tools, e.g. *tar,* is prohibitively slow.

*dd* was written by Ken Thompson and Dennis Ritchie as part of Bell Labs' original Unix operating system version 5. [6,7]. It was propagated as one of the core utilities to all later versions of Unix and has been included in Linux since its inception. *dd* uses *open*(), *lseek*(), and *read*() to access files [8]. The time required to access any data block within a large file using lseek() is independent of the location of the block within the file. Furthermore, the readout of a continuous set of records is handled as a streaming operation at the maximum speed provided by the file system. Hence *dd* enables high speed utilization of large files as random access memory.

The general usefulness of this capability is presented in solutions to (a) the page image problem mentioned above, (b) a second more complex genealogical database search problem solved also in a shared web hosting environment, and (c) a neuroimaging research problem solved in a shared supercomputing environment.

A Random Access Concatenated Library (RacLib) is a single file containing a collection of equal-length records. A record set is a contiguous sequence of one or more record(s). A second index file contains a one-line entry for each record set. At minimum this line contains a label for the corresponding record set, the starting record number, and the number of records in the record set.

Each index entry may contain additional record set information. In some cases, e.g. the Social Security Death Index (SSDI) problem (see below), the index file has sufficient intrinsic structure to enable organizing it also as a RacLib in which case its length may be unlimited. If not, the index file must be searched in a serial fashion to find the entry for the required RacLib record. In that case, the length of the index is best limited to 100,000 lines or less.

The file structures were designed for easy and efficient access using CGI script languages. Access to the files uses standard Linux tools, e.g. *grep* and *dd*, hence comparable performance may be expected with any scripting language.

**Image Delivery**

DonsList.net delivers human readable page images as shown in Figure 1 from genealogical resources, e.g. census listings, military rolls, yearbooks, city and phone directories. The website includes 3,500,000+ page images from 8000+ titles. Each jpeg image is about 250 Kbytes. For delivery at the request of a web browser user, an image must be extracted from the RacLib where it is stored and placed in a website-accessible cache directory as a jpg file. Management of the cache is detailed below.

The website is limited by the shared hosting service to 50,000 files. In addition, the response time of the website is limited by the load placed on the shared server by hundreds of other websites housed on the same server. One may compare the display capabilities and response time of DonsList [9] with that provided by the Internet Archive [10], a well-funded library which houses 35+ million volumes totaling billions of pages and which utilizes dedicated storage and compute servers.

In order to manage the file count, all of the page images from one or more of the titles are concatenated into a single file, a RacLib. The length of each page image is padded with null characters to an integer number of 1024-byte records. Note that jpeg images are compressed, hence no additional compression need be applied to the RacLib to minimize its size.

Each line in the corresponding index file contains the book title, the page number within it, the starting record number and the number of records for the page image, e.g. "TallyHo1965 0404 4050730 348" This example is page 404 from the 1965 Florida State University Yearbook. The image begins at record number 4050730 and contains 348 1024-byte records. TallyHo1965 is one of the series of 1901-2007 Florida College for Women/FSU yearbooks housed on the website (Figure 1, upper right frame), 72 in total, all placed in a single RacLib and indexed in a single index file. The total number of page images in this RacLib and hence index lines is 21,988.

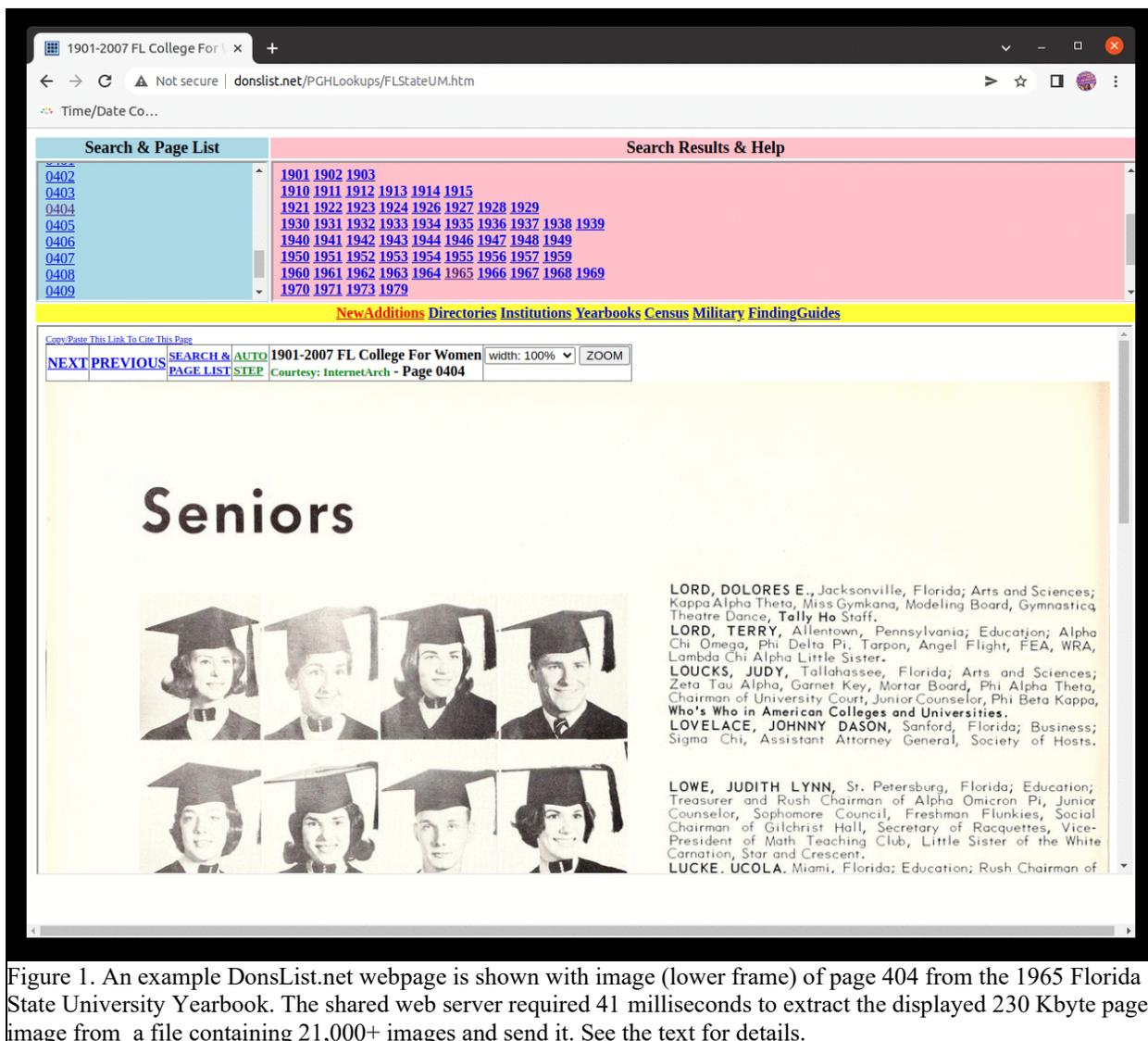

Figure 1. An example DonsList.net webpage is shown with image (lower frame) of page 404 from the 1965 Florida State University Yearbook. The shared web server required 41 milliseconds to extract the displayed 230 Kbyte page image from a file containing 21,000+ images and send it. See the text for details.

Fetch time from the RacLib to the image cache for the image shown in Figure 1 was 41 milliseconds (msec) The index line number for this page image is 15,582, i.e. 41 msec included the time required for a serial search through 75% of the entries in the index file. The sequence of operations performed by each CGI instance in response to a webpage request are shown in Figures 2 and 3 and detailed as follows.

- Parse the request and capture the requested book title and page number.
- If the requested page image is present in the cache directory, generate the requested webpage with link to the cached image, send, and then terminate the current CGI instance.
- If the requested page image is not present,
  - fetch the record number and count for the requested page image from the corresponding index file. Note that this fetch requires a serial search of the index file using Linux utility *grep*. The search matches the book title and page number, i.e. the lines in the index file are content addressable.

- Fetch the corresponding page image from the corresponding RacLib and write it to the cache directory. The RacLib is utilized at this step as random access memory by Linux utility *dd*.
- Generate the requested webpage with link to the cached image, send, and then terminate the current CGI instance.

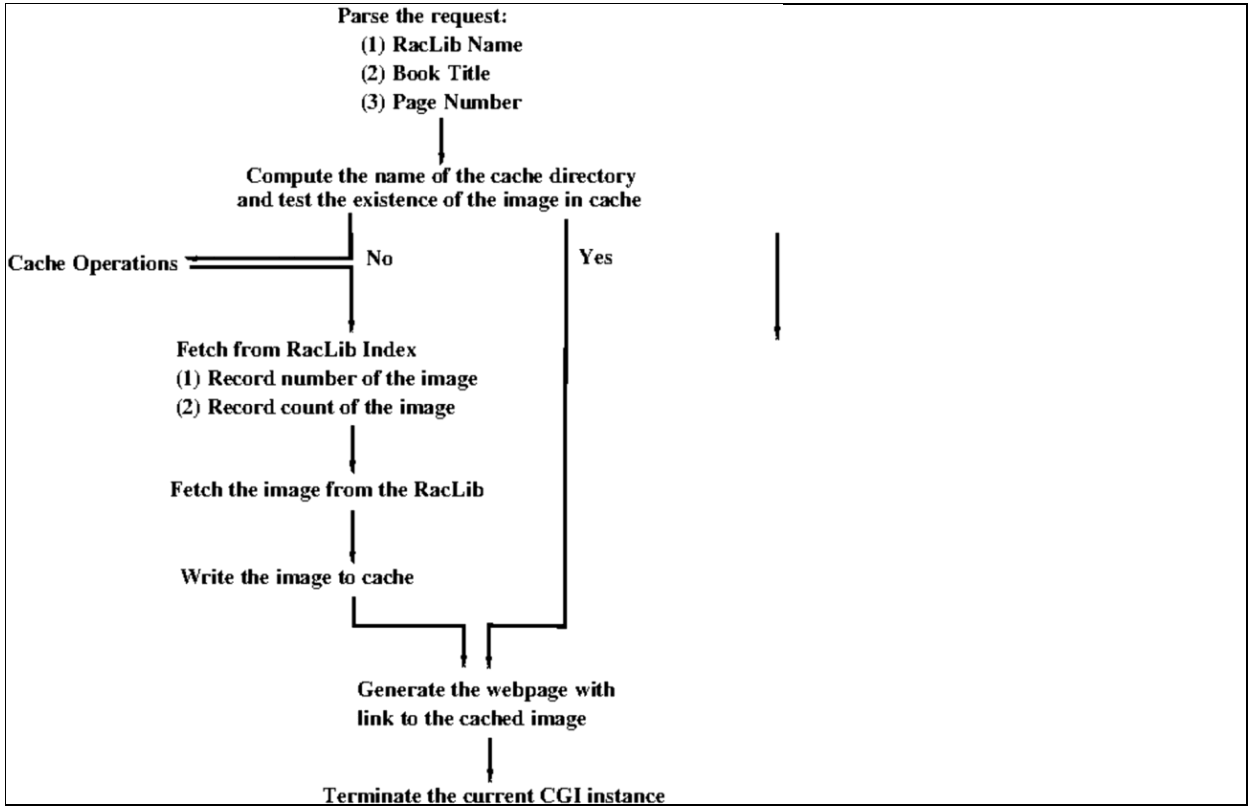

Figure 2. Sequence of operations performed by each CGI instance in response to a webpage request. See the text for more detail. The "Cache Operations" indicated in the figure are shown in Figure 3.

*dd's* use of *lseek()* enables extraction of records with equivalent latency independent of the location within the file. Here are spot checks on the times required to fetch 230 1024-byte records from near the beginning and end of a 6-Gbyte file. Note that the test was run on the later part of the file first.

    36 msec: time dd if=FLStateU.RacLib bs=1024 count=230 skip=6069853 > /dev/null [11,12]
    41 msec: time dd if=FLStateU.RacLib bs=1024 count=230 skip=9398 > /dev/null

The file system is used as content-addressable memory since (a) the book title is present in the name of the corresponding RacLib and index files and (b) the page image name and page number are present in the name of the corresponding page image in the cache directory. This expedient markedly reduced the time and effort required to develop the CGI scripts.

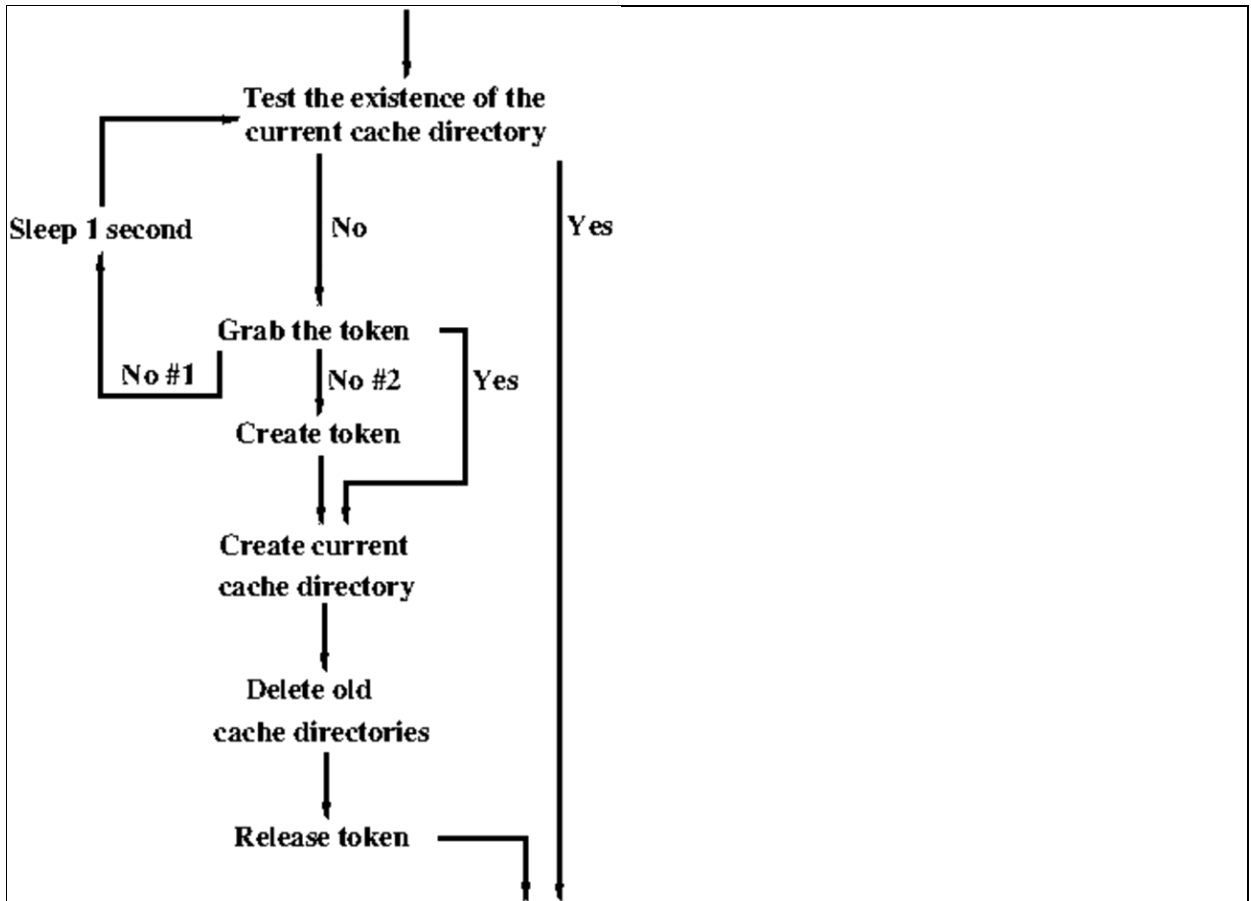

Figure 3. Sequence of cache management operations performed by each CGI instance in response to a webpage request. This figure shows the "Cache Operations" indicated near the top of figure 2. See the text for more detail.

As mentioned above, for delivery at the request of a web browser user, an image must be extracted from the RacLib where it is stored and placed in a cache directory as a jpg file. Management of the cache is handled as follows (see Figure 3). The current cache directory with page image files persists for 2,000 seconds. The directory name is selected using Linux utilities: *date +%s | sed -e "s/…$//"*. *date +%s* computes the number of 1000-second intervals that have elapsed since January 1, 1970 and *sed -e "s/…$//"* removes the last three digits.

Each CGI instance computes the directory name and checks for its existence. If an instance fails to find the cache directory, it attempts to grab the *token* file. This is accomplished by renaming the token file using the CGI instance's unique process ID in the new temporary name. Releasing the token is accomplished by renaming the file to the original name. A *token* is used to minimize the chance of a race condition in which two or more CGI instances attempt to perform the same file system manipulations at the same time. The instance which grabs the *token* creates the current cache directory, deletes any cache directories which are more than 2000 seconds old, releases the token, and continues. The CGI instances which fail to grab the token sleep for one second and check for the current directory again. Any CGI instance which fails a second time creates the token and grabs it. This ensures that the system will not hang indefinitely if a CGI instance which owns the token fails to release it. Note that this approach obviates the need for a daemon process which handles cache housekeeping.

This system has been in place for 12+ years without a failure. During that period, it has delivered 11,000,000+ page images, 9,800,000+ fetched from a RacLib and 1,400,000 previously cached images. Figure 4 shows histograms of the response time of the system for both RacLib fetches (upper panel) and cache hits (lower panel). The cache hit *mean* response time is 24 msec faster than the RacLib fetch *mean* response time. The RacLib fetch response times from 2012-2013 are about 15 msec faster than those from 2020-2022, perhaps due to heavier load on the shared server and/or changes in the server's disk subsystems over the years.

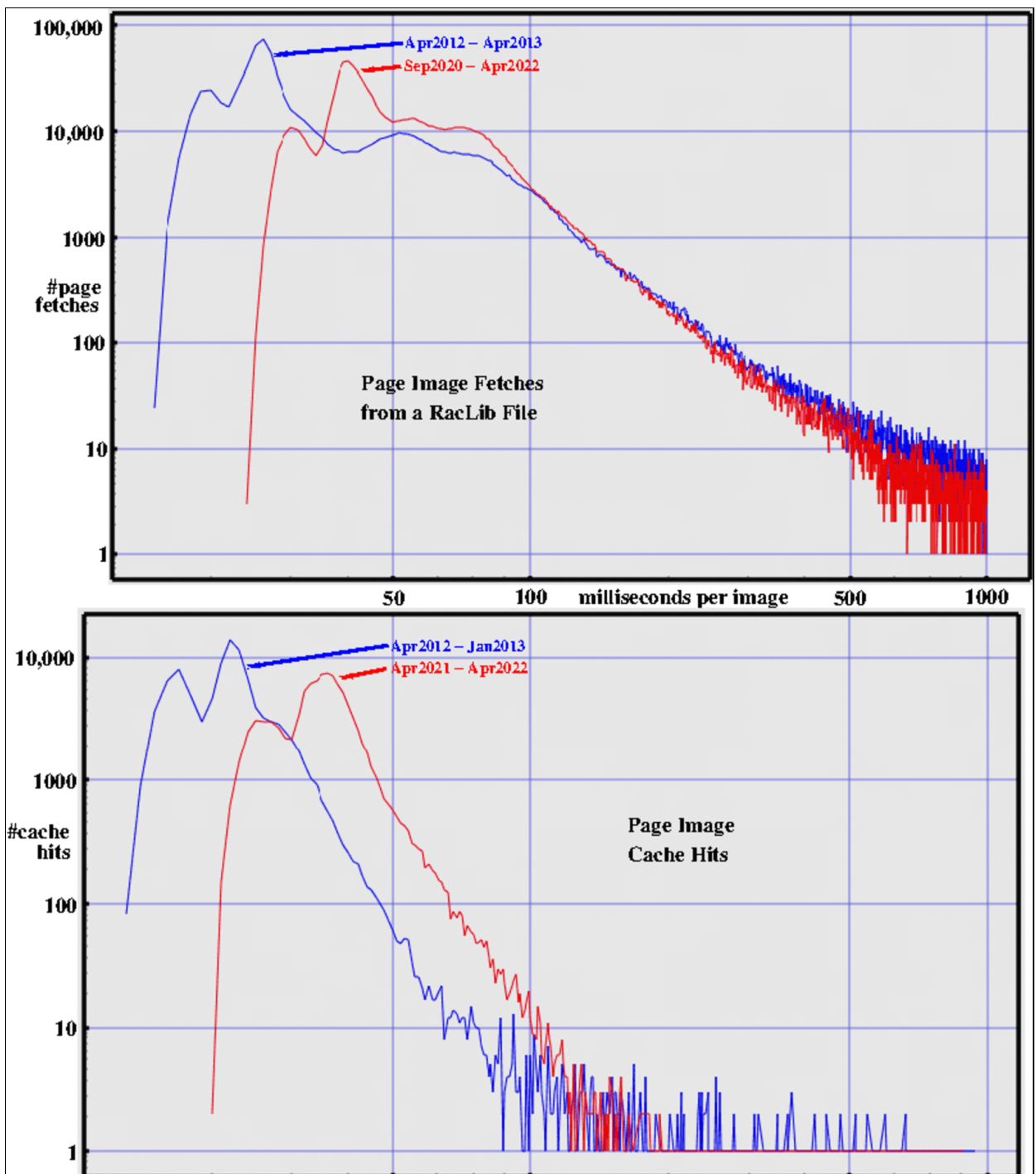

Figure 4. The page image RacLib system has been in place since 2012 and has delivered 11,000,000+ page images to date: 9,800,000+ fetched from a RacLib and 1,400,000 previously cached images. Histograms of the response time of the system for both RacLib fetches (upper panel) and cache hits (lower panel) are shown for the first (blue) and last (red) 1,000,000 RacLib fetches and for the first and last 100,000 cache hits. Log-Log axes are used to enable appreciation of the relatively rare instances with prolonged response times. See text for more details.

**Social Security Death Index (SSDI)**

DonsList.net also houses a copy of the SSDI. This is a database containing 86,806,853 64-byte records, i.e. 5.5 GBytes. This is large enough to significantly degrade the performance of a relational database management system [13]. The SSDI lists every deceased person with a US Social Security number for whom the Social Security death benefit has been filed since the inception of the system in 1936. This version is current to 2011. Each decedent is listed with given name and surname, Social Security number, birth date, and death date. **Figure 5** shows a sample search with results from DonsList. Note that the birth date and death month and year of the first entry in the results match those of US Attorney General Robert F. Kennedy, and that the search required 39 msec. One may compare the search capabilities and response time of DonsList [14] with those provided by large commercial libraries [15-17], all of which utilize dedicated storage and compute servers.

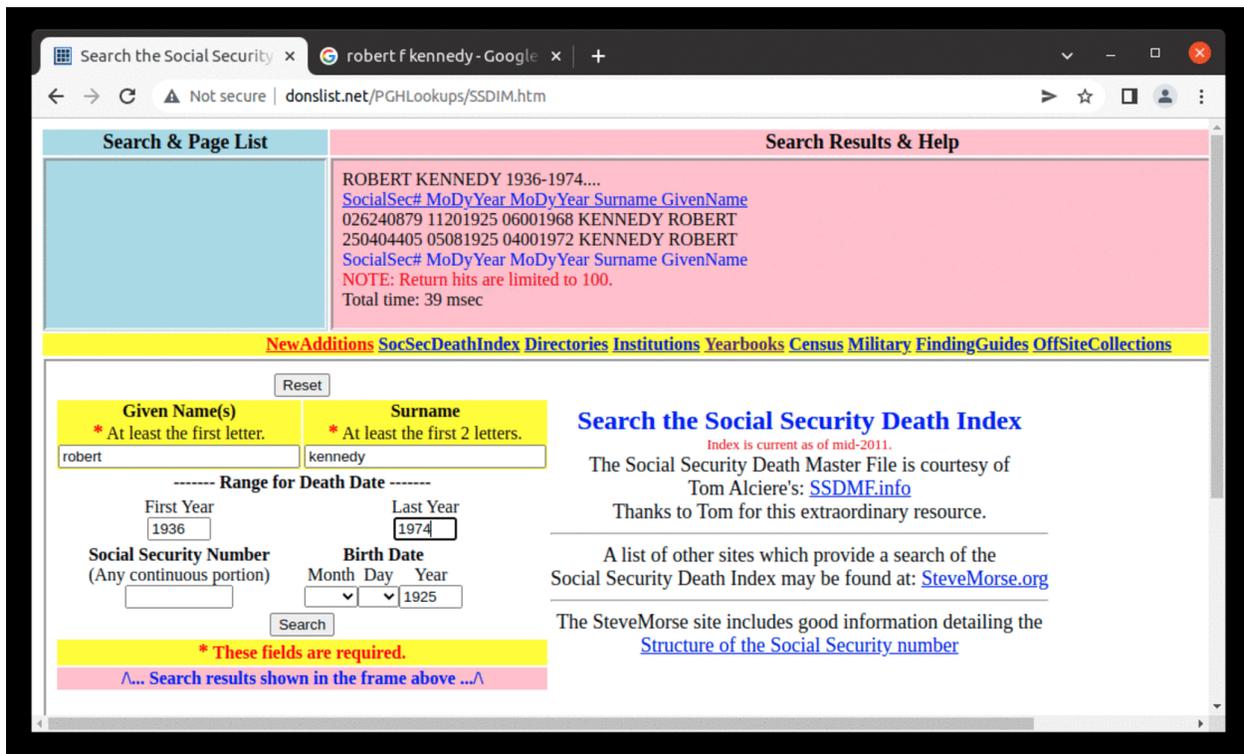

Figure 5. An example SSDI search (lower frame) with results (upper right frame) is shown for "Robert Kennedy," death years ranging from 1936 to 1974, and birth year 1925. The shared web server required 39 msec to extract the results, and send them. See the text for details.

The SSDI dataset possesses an intrinsic structure which is utilized to organize the RacLib and which enables organizing the index file as a RacLib also. The records in the dataset are concatenated in the RacLib using a sort on the first letter of the surname, then the second letter of the surname, and then the first letter of the given name. All of the records with the same three key letters therefore fall together into one of $26^3$ record groups.

Each index line contains the starting record number and the number of records in one of these 17,576 groups. The location of each index entry is calculated using the numbers corresponding to the three key letters, i.e. the index is also a RacLib and randomly rather than serially accessed.

For the search shown in Figure 5 on Robert Kennedy, the key letters are K(10), then E(4), then R(17). The number is the enumeration for each letter with A, the first letter of the alphabet, being 0 (zero). Using these numbers, the location of the corresponding index record is $(17_R) + (4_E \times 26) + (10_K \times 26^2) = 6881$.

Here is the sequence of operations performed by a CGI instance in response to an SSDI search request (Figure 6).
- Parse the request and capture the names and any other information provided for the search (see Figure 5).
- Compute the record number within the index RacLib using the three key letters and fetch the index record. This is a random access fetch using Linux utility *dd*.
- Use the record number and count contained in the index record to fetch the set of SSDI records from the SSDI RacLib using Linux utility *dd*.
- Search the SSDI record set using all information provided for the search, send the results, and terminate the current CGI instance. Note that this is a serial search using Linux utility *grep* of all the SSDI records which match the three key letters.
- Generate the requested webpage, send, and then terminate the current CGI instance.

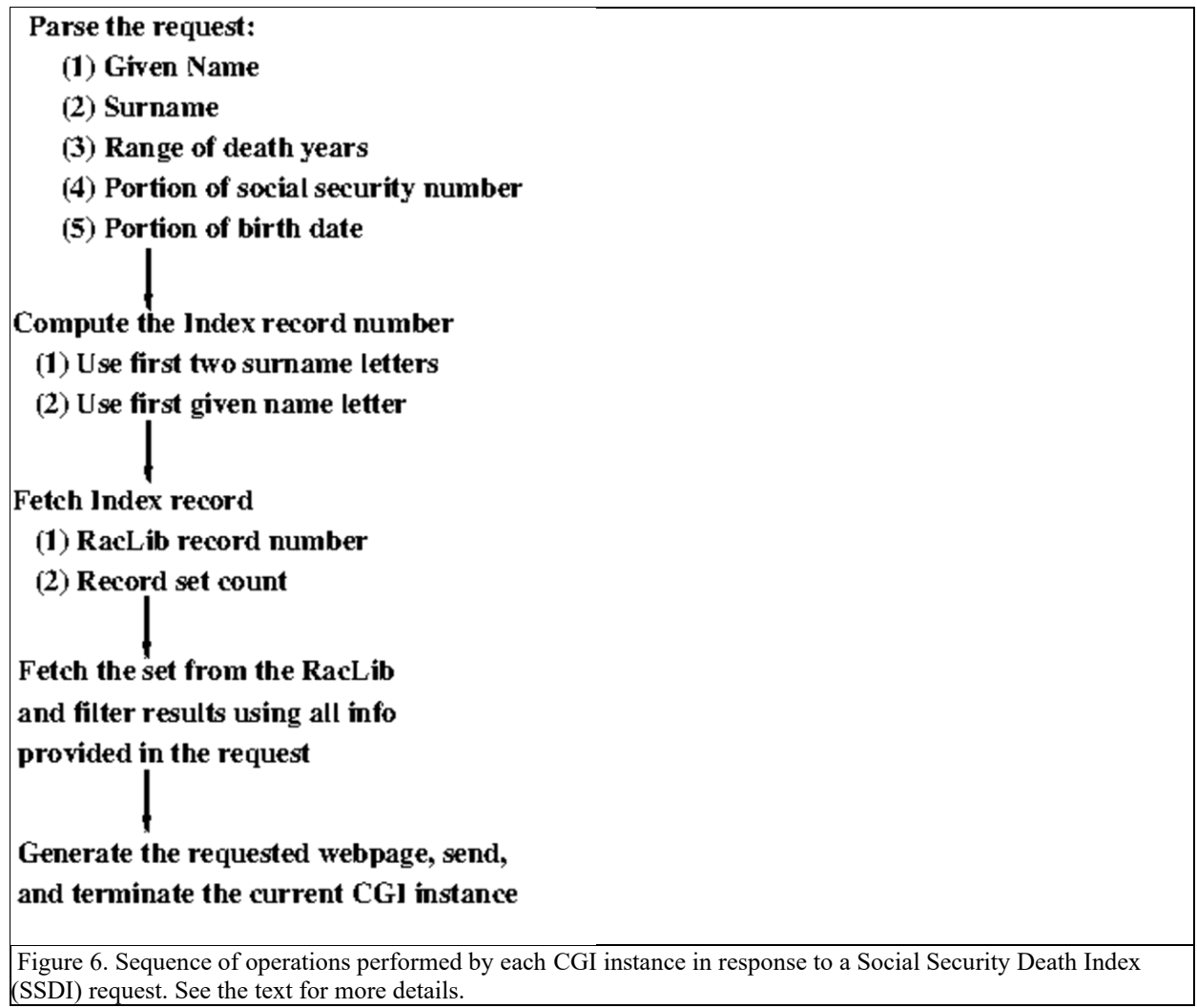

Figure 6. Sequence of operations performed by each CGI instance in response to a Social Security Death Index (SSDI) request. See the text for more details.

129 of the SSDI record sets corresponding to a single three-letter index contain more than 100,000 64-byte records, i.e. more than 6,400,000 bytes each. For example, the set of records for key letters "JO J" contains 142,989 records since it includes given names James, John, and Joseph and surnames Johnson and Jones. Over the decade that the SSDI RacLib system has been in place, that search ($n = 148$) required a *mean* of 149 msec with *standard deviation* of 150 msec. Figure 7 shows a histogram of the response time of the system. The searches which required more than 150 msec were likely primarily due to searches of these large record sets.

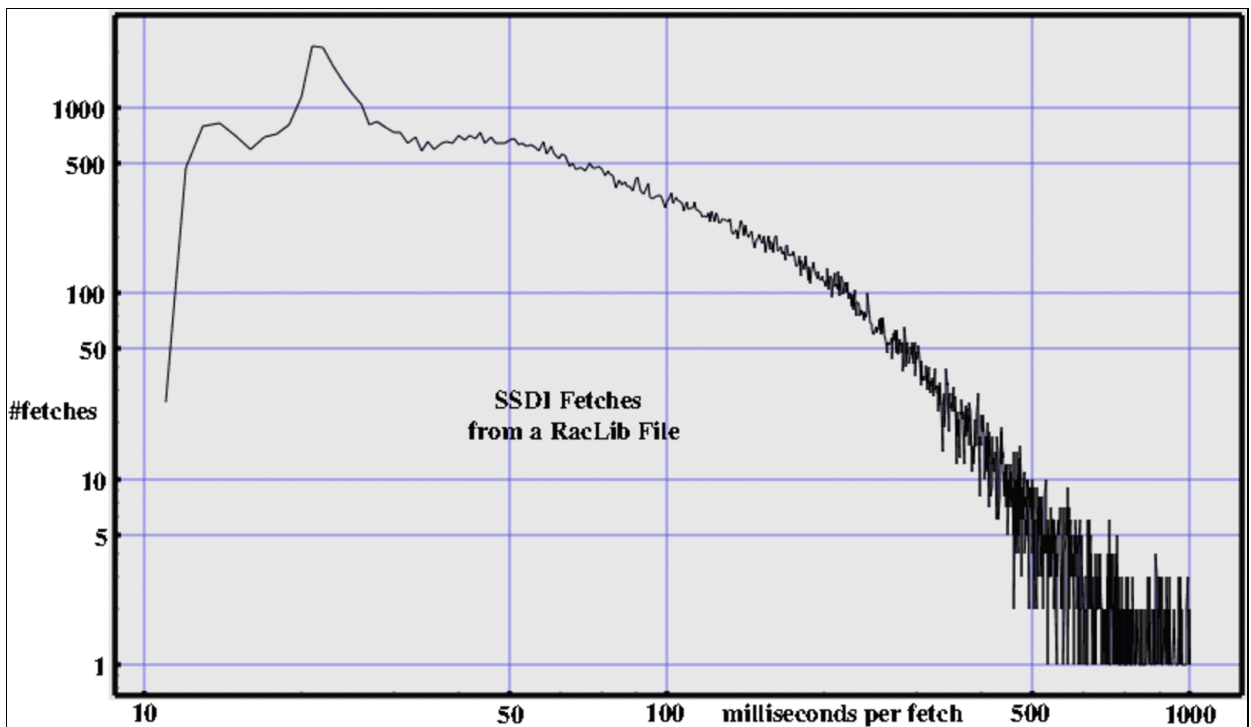

Figure 7. The SSDI RacLib system has been in place since 2012 and has delivered 100,000+ searches to date. A histogram of the response time of the system is shown. Log-Log axes are used to enable appreciation of the relatively rare instances with prolonged response times. See text for more details.

## Neuroimaging

Anatomically stereotypic brain regions are automatically identifiable with millimeter (mm) accuracy using high resolution magnetic resonance imaging (MRI) and computerized brain segmentation tools, e.g. freesurfer [18] and tracula [19]. With these tools, the stereotypic region which contains each cubic millimeter (mm$^3$) of brain tissue is identified along with the xyz coordinates of the tissue in the MRI coordinate system. The xyz coordinate list for each region is typically stored in a separate file. For work carried out in our lab [20], the lists by regions were compiled into a single RacLib for each of 800+ human subjects.

The xyz coordinates for each 1 mm$^3$ volume are represented by a single string, e.g. n41p12n35 represents (-41,12,-35). The individual brain is divided into cubic centimeters (cm$^3$), each containing up to 1000 mm$^3$ xyz coordinates, e.g. n4_xp1_yn3_z represents (-4[0-9],1[0-9],-3[0-9]). Within the RacLib, all of the cm3 volumes which overlap a single region are grouped together; all the mm$^3$ xyz coordinates within a cm$^3$ volume are grouped together.

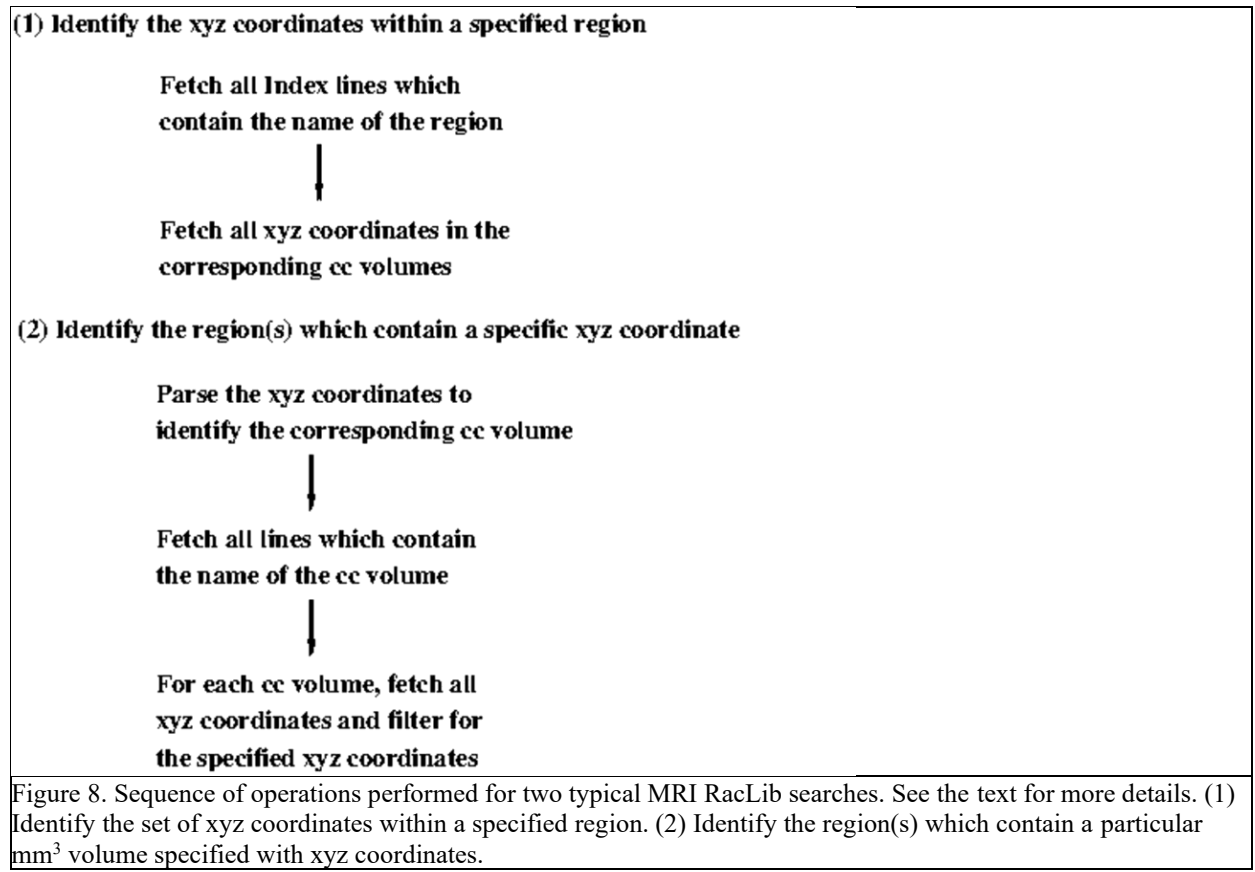

Figure 8. Sequence of operations performed for two typical MRI RacLib searches. See the text for more details. (1) Identify the set of xyz coordinates within a specified region. (2) Identify the region(s) which contain a particular mm$^3$ volume specified with xyz coordinates.

Each entry in the index file contains the name of a region, the name of a cm$^3$ volume which overlaps the region, the starting record number, and the number of mm$^3$ xyz coordinates within the cm$^3$ volume. Here is an example index entry:
    L_ctx_middletemporal n4_xp1_yn3_z 85652 181

The structure of the RacLib and its index enables easy and rapid searches of many types (Figure 8). The organization of the RacLib into record sets for each cm$^3$ volume coupled with the RacLib's random access capability are particularly effective in reducing the load required to identify the region which contains a specific xyz coordinate, i.e. (2) in Figure 8. Each record set contains a maximum of 1000 records which must be searched whereas the number of 1-mm$^3$ volumes within a region ranges up to 60,000. Use of RacLibs containing mm-resolution neuroimaging data as described was integral to developing the results shown in figure 9 [20].

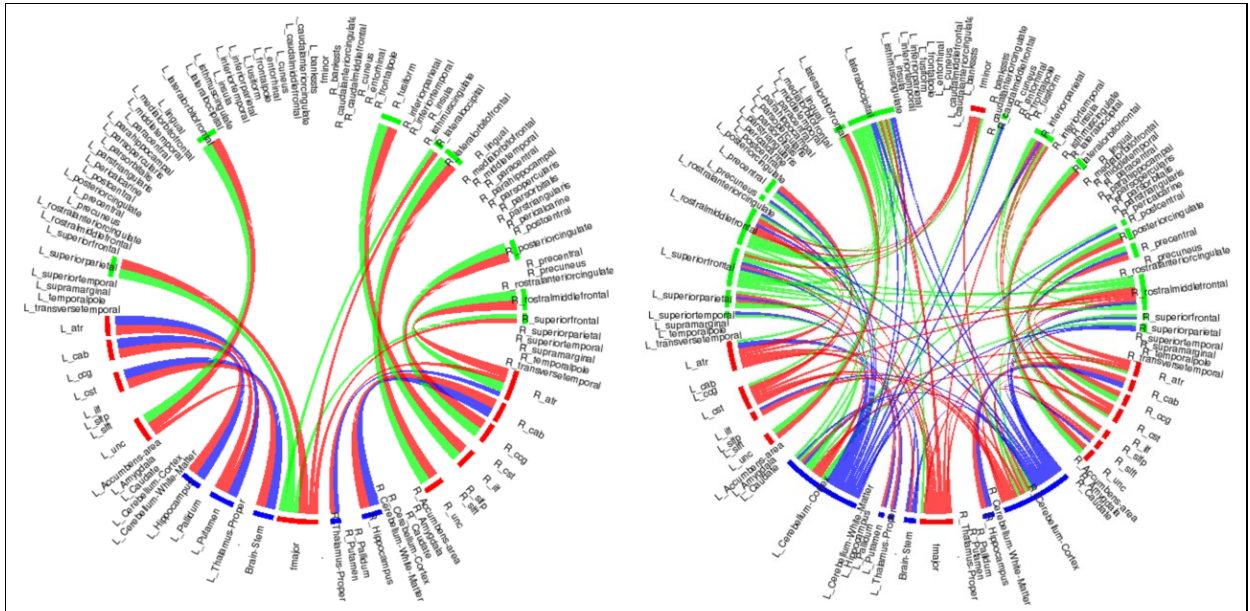

Figure 9. Pairs of brain regions in this individual which are significantly coupled (p < 10$^{-8}$) are connected by chords. Green/red/blue chords connect cortical/subcortical/deep white matter regions. Coupling strength was measured by counting simultaneous appearance of electrical currents extracted from magnetoencephalographic recordings [20]. Relative coupling strength is proportional to the circumferential length of the bars on the circle just inside the region names. The left panel show results from twelve minutes of resting with eyes open in dim light. The right panel shows results from twelve minutes of cognitive task performance.

The Figure shows pairs of brain regions in an individual (human) which are functionally coupled to each other during rest (left panel) and cognitive task performance (right panel). The functional measure is consistent occurrence of transient neuroelectric currents in two regions at once, p < 10-8 for each pair of regions. The currents were extracted from magnetoencephalographic (MEG) recordings. Participants' approvals for participation were obtained using written informed consent approved by the University of Pittsburgh IRB (PRO13070121). Note that fmajor is the only region which is connected from one side to the other during rest. fmajor is the posterior portion of the corpus callosum, a deep white matter tract made up of fibers which connect the two sides of the brain. In contrast, florid left/right regional coupling is seen during task performance. Note also that both left and right motor areas demonstrate coupling to other regions during task performance, e.g. precentral and Cerebellum, but not during rest.

## Summary

RacLib databases accessed using Unix/Linux utility *dd* enable (a) scalable aggregation of up to 100,000 jpeg or other files into a single library file and (b) very rapid record fetches with equivalent latency regardless of record location within the library. Use of a human-readable ASCII index file simplifies and expedites program development. These properties have enabled the development and deployment of a fast and reliable website on an inexpensive shared web server which houses and provides short-latency access to 3,500,000+ page images and a database containing 86,000,000+ records. Use of RacLib files will scale to much larger library websites. RacLib databases have also proven useful in research application deployed on shared cluster computing and supercomputing resources. It is noteworthy that technology which has not appreciably changed for 48 years enables expedited programming and low-latency performance with limited resources.